\documentclass{article}

\usepackage{arxiv}

\usepackage[utf8]{inputenc} 
\usepackage[T1]{fontenc}    
\usepackage{hyperref}       
\usepackage{url}            
\usepackage{booktabs}       
\usepackage{amsfonts}       
\usepackage{nicefrac}       
\usepackage{microtype}      
\usepackage{lipsum}
\usepackage{graphicx}
\graphicspath{ {./images/} }
\usepackage{multirow}
\usepackage{lineno}
\usepackage{amsmath}
\usepackage{amssymb}

\title{MMBAttn: \textbf{M}ax-\textbf{M}ean and \textbf{B}it-wise \textbf{Att}e\textbf{n}tion for CTR Prediction}

\author{
 Hasan Saribas \\
  AI Enablement\\
  Huawei Türkiye R\&D Center\\
  Istanbul, Turkey \\
  \texttt{hasan.saribas1@huawei.com} \\
   \And
 Cagri Yesil \\
  AI Enablement\\
  Huawei Türkiye R\&D Center\\
  Istanbul, Turkey \\
  \texttt{cagri.yesil1@huawei.com} \\
  \And
 Serdarcan Dilbaz \\
  AI Enablement\\
  Huawei Türkiye R\&D Center\\
  Istanbul, Turkey \\
  \texttt{serdarcan.dilbaz@huawei.com} \\
   \And
 Halit Orenbas \\
  AI Enablement\\
  Huawei Türkiye R\&D Center\\
  Istanbul, Turkey \\
  \texttt{halit.orenbas@huawei.com} \\
}

\begin{document}
\maketitle
\begin{abstract}
With the increasing complexity and scale of click-through rate (CTR) prediction tasks in online advertising and recommendation systems, accurately estimating the importance of features has become a critical aspect of developing effective models. In this paper, we propose an attention-based approach that leverages max and mean pooling operations, along with a bit-wise attention mechanism, to enhance feature importance estimation in CTR prediction. Traditionally, pooling operations such as max and mean pooling have been widely used to extract relevant information from features. However, these operations can lead to information loss and hinder the accurate determination of feature importance. To address this challenge, we propose a novel attention architecture that utilizes a bit-based attention structure that emphasizes the relationships between all bits in features, together with maximum and mean pooling. By considering the fine-grained interactions at the bit level, our method aims to capture intricate patterns and dependencies that might be overlooked by traditional pooling operations. To examine the effectiveness of the proposed method, experiments have been conducted on three public datasets. The experiments demonstrated that the proposed method significantly improves the performance of the base models to achieve state-of-the-art results.
\end{abstract}


\section{Introduction}
Click-through rate (CTR) prediction is a critical problem in web search, recommendation systems and online advertisement displaying.
CTR prediction is important because it helps service providers attract and engage their users by ranking a small number of items from a large number of candidates.
Accurate CTR prediction allows for better optimization of advertising campaigns, improved user targeting, and enhanced revenue generation.
Improving the accuracy of advertising CTR can not only make the user experience better but also give more benefits to advertising platforms and advertisers.
Since many platforms rely primarily on advertisement revenues as their primary mode of income, the user base for advertisements is very large. Traditional approaches to CTR prediction have relied on statistical methods and hand-engineered features. While these methods have achieved moderate success, they often struggle to capture the complex and nonlinear relationships inherent in user behavior and contextual factors. Moreover, they require extensive manual feature engineering, making them less scalable and adaptable to dynamic advertising landscapes. Machine Learning (ML) techniques have been able to automatically discover relevant interactions for CTR but with rapid innovations in deep learning (DL) architectures, DL models have been able to better learn high-order feature interactions in sparse spaces. In recent years, deep learning has emerged as a powerful paradigm for solving complex prediction tasks by automatically learning hierarchical representations from raw data. This approach has shown remarkable success in various domains, such as computer vision and natural language processing. Consequently, researchers have directed their attention towards developing deep learning models for CTR prediction, aiming to leverage their capacity to capture intricate patterns and dependencies within large-scale datasets. Early approaches in CTR prediction leveraged logistic regression but these approaches lacked the ability capture higher-order interactions \cite{logistic2013trenches, logistic2007}. Later, the factorization machine (FM) structure gained popularity since FM projects sparse features onto low-dimensional space and learns feature interactions via inner products \cite{fm}. Over time, as deep learning approaches improved, a plethora of deep learning models such as Wide\&Deep, DeepFM, DCN, AutoInt, FiBiNET have seen widespread use for CTR prediction \cite{wide&deep, guo2017deepfm, deep&cross, song2019autoint, huang2019fibinet}. The attention mechanism has proven very powerful in deep learning and has been utilized heavily in state-of-the-art models in many domains \cite{attn2021review}. Attention has been applied for CTR prediction as well \cite{huang2019fibinet, zhang2022fibinet++, song2019autoint}.
In this article, we present a novel deep learning model where \textbf{M}ax-\textbf{M}ean and \textbf{B}it-wise \textbf{Att}e\textbf{n}tion (MMBAttn) are leveraged. Our model addresses the limitations of traditional methods and builds upon the state-of-the-art deep learning techniques by efficiently leveraging the attention mechanism with the lightweight maximum and mean operations.
The main contributions of this paper are summarized as follows:
\begin{itemize}
  \item We propose a new attention architecture (MMBAttn) for CTR prediction to emphasize important features.
  \item The proposed method is designed as a plug-and-play module and is therefore model-agnostic. It can be easily applied to any CTR prediction model.
  \item We performed extensive experiments on three public datasets, namely Criteo, Avazu, and Frappe. Additionally, we conducted ablation studies on Criteo and Avazu datasets to evaluate the performance of each component proposed by our method. The results indicate that the inclusion of each additional component leads to improved performance.
\end{itemize}

\begin{figure}[t]
\begin{center}
\includegraphics[width=\textwidth]{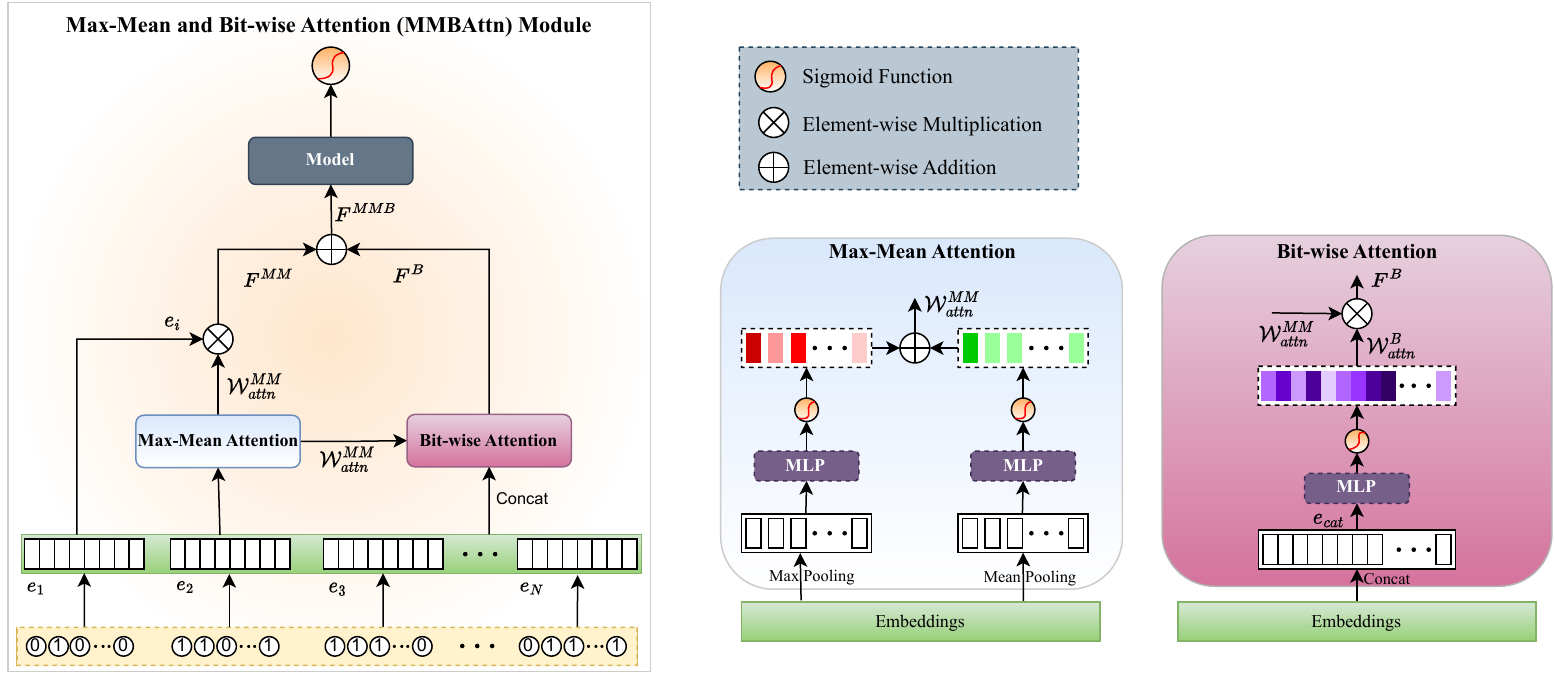}
\caption{The general architecture of the proposed attention mechanism.}\label{fig:architecture}
\end{center}
\end{figure}

\subsection{Related Works}

\subsubsection{Click-through Rate Prediction}
Logistic regression (LR) was one of the early proposed models for CTR prediction and it has proven to be a strong baseline \cite{logistic2013trenches, logistic2007}. Nonetheless, LR cannot accurately learn nonlinear interactions, so factorization machines (FM) were introduced. Opposed to LR, FM embeds the features to a dense so that the feature interactions can be modelled via inner products \cite{fm}. Multilayer perceptron was one of the earlier deep learning models to be used in CTR prediction \cite{covington2016deep}. Later, Wide\&Deep model improved on the multilayer perceptron structure by combining a wide network with a deep network to leverage the advantages of both approaches \cite{wide&deep}. DeepFM model built on the FM and Wide\&Deep models to learn second-order feature interactions \cite{guo2017deepfm}. Deep\&Cross Network (DCN) learns to model an even higher order of interactions to perform better CTR prediction \cite{deep&cross}.

\subsubsection{Attention Networks}
The attention mechanism was first proposed for neural machine translation where attention overcomes the information bottleneck created by recurrent neural network architectures \cite{bahdanau2014neural}. The seminal work from \cite{vaswani2017attention} showed that the attention mechanism can be used on its own to learn powerful representations from sequential data to tackle problems ranging from text summarization \cite{el2021automatic} to question answering \cite{yu2019deep}.  The success of attention in natural language processing led to the widespread usage of attention in other domains. Many of the seminal works in computer vision utilized the attention structure to obtain architectures that learn better representations. Attention was initially added to existing architectures where commonly used vision blocks such as convolution and pooling were combined with attention \cite{wang2020eca, hu2018squeeze, woo2018cbam}. Models that primarily rely on attention were able to achieve state-of-the-art results in the vision domain \cite{khan2022transformers, dosovitskiy2021image}. Based on this widespread success, attention mechanisms have also been used for CTR prediction.

\subsubsection{Attention Networks for CTR}
In this paper, we propose a new attention architecture (MMBAttn) for CTR prediction to emphasize important features. Attention mechanisms have shown great promise in improving the performance of recommendation systems by allowing them to attend to the most informative user-item interactions. For instance, the Attentional Factorization Machine (AFM) model \cite{xiao2017attentional} applies the attention mechanism to the intermediate output of the factorization machine layer to focus on the pairwise interaction of feature embeddings. The Deep Interest Evolution Network (DIEN) and Deep Session Interest Network (DSIN) models \cite{zhou2019deep, feng2019deep} do not apply attention to all the features. They utilize self-attention mechanisms to only sequential features to capture user behavior changes. Moreover, there are models that apply attention mechanisms not just to sequential features but also to categorical and numeric feature embeddings or feature interactions. For instance, the Autoint model \cite{song2019autoint} uses a multi-head self-attentive neural network with residual connections to explicitly model feature interactions in low-dimensional space. The InterHAt model \cite{li2020interpretable} applies a transformer with multi-head self-attention to input embeddings to capture interactive relations between features without using the Multi-Layer Perceptron (MLP) structure. After the multi-head self-attention layer, a hierarchical attention mechanism is utilized to benefit from high-order interactions. In contrast to Autoint and InterHAt, the Gated Attention Transformer (GAT) model \cite{long2021efficient} uses vanilla attention mechanisms and Gated Attention Transformer instead of self-attention to overcome the drawbacks of Autoint and InterHAt.
On the other hand, the Fibinet and Fibinet++ models \cite{huang2019fibinet, zhang2022fibinet++} learn the importance of features with the Squeeze-Excitation (SENET) and SENET+ mechanisms which are initially used in computer vision domain \cite{wang2020eca}. The SENET and SENET+ mechanisms consist of three components: squeeze, excitation, and re-weighting. In the squeeze part of SENET, the i-th feature is summarized with mean or max pooling, and the squeezed value represents the global information about the i-th feature embedding. In the excitation part, a weight is learned for each feature embedding by applying an MLP to the squeezed value vector created in the squeeze part. In the re-weight part, new feature embeddings are created by multiplying the learned weights and the original feature embeddings. The SENET+ updates the squeeze parts of SENET by segmenting each feature embedding into $g$ groups. Then, both the max and mean values of each group are calculated separately, and the resulting vector is forwarded to the excitation part. Finally, Fibinet and Fibinet++ combine the output of SENET and SENET+ layers with a bilinear interaction layer and forward it to MLP layers for prediction.
The FAT-DeepFFM model \cite{zhang2019fat} introduces the Compose-Excitation network (CENet) field attention mechanism, which is an enhanced version of SENET to highlight feature importance. CENet combines SENET with the DeepFFM model, which is based on the Field-aware factorization machines (NFM) \cite{juan2017field}. This study also applies CENet on features before explicit feature interaction and on cross features after explicit feature interaction for comparison. Unlike SENET, CENet carries out the pooling operation with 1x1 convolution instead of max and mean pooling in the excitation part. Then, two fully connected (FC) layers are used for the excitation part, as in SENET.
The ECANFM model \cite{shi2021ctr} proposes the ECANET concept, which is based on the ECA module used in the computer vision domain, and combines it with the NFM model \cite{he2017neural}. The ECANFM model aims to improve feature embeddings by dynamically learning the importance of feature embeddings. ECANET uses mean pooling for the squeeze part. But in the excitation part, it uses one-dimensional convolution instead of fully connected layers.
Lastly Masknet model \cite{wang2021masknet} uses a kind of attention mechanism that learns weights for value in feature embeddings to highlight the important parts. Unlike SENET and its variants, Masknet does not use a squeeze module. It normalizes the feature embeddings and directly applies fully connected layers to the normalized features to learn the weights.
This study presents a novel approach to feature importance estimation that differs from existing works. Specifically, we propose an attention structure that combines max and mean pooling operations with multiple MLPs to obtain feature importances. To the best of our knowledge, there are only two studies that use max and mean pooling together including Fibinet++ \cite{zhang2022fibinet++} and CBAM \cite{woo2018cbam}. In Fibinet++ max and mean pooling are used together in a group-wise structure that employs shared MLPs in the excitation step. CBAM uses max and mean pooling operations with shared MLPs in the vision domain. In contrast, our approach uses Max-Mean pooling operations with separate MLPs. Unlike using shared MLPs employed in previous studies, our approach uses Max-Mean pooling operations with two distinct MLPs to emphasize feature importance that comes from max and mean pooling branches separately. Unlike previous methods that rely solely on max and mean pooling operations, our approach also employs a bit-wise structure that captures attention between all bits using a distinct MLP. This allows us to utilize information that may be lost during pooling operations. Importantly, our proposed attention structure integrates all these components to provide a comprehensive and effective feature importance estimation method. Overall, our study contributes to the field by introducing a novel approach that improves upon existing methods and provides a more accurate and comprehensive understanding of feature importance.

\section{Method}
In this study, an attention-based method has been proposed to better highlight the importance of features. Fig.~\ref{fig:architecture} illustrates the attention-based method. In the proposed approach, mean and max pooling operations are applied to the features to obtain feature importance. However, these pooling operations may result in information loss. To address this, a bit-wise attention structure is additionally utilized. With the bit-wise structure, we can emphasize the relationships between all bits across all features and achieve an architecture that highlights these relationships. The proposed attention-based models are designed as a plug-and-play module, hence they can be applied to all existing CTR prediction models.

\noindent\textbf{Embedding layer.} CTR prediction models often deal with high-dimensional and sparse input features, such as categorical variables representing user information, item details, and contextual factors. To effectively capture the relationships and interactions among these features, feature embedding layers are utilized. The embedding of a categorical variable $x_i$ in an embedding layer can be calculated using the following equation:

\begin{equation}
e_i = \boldsymbol{W_e}x_i,
\end{equation}

\noindent where $x_i \in R^{1\times d}$ is the one-hot representation of the categorical variable $x$ at index $i$. It is a binary vector of size $V$, where $V$ is the vocabulary size (number of unique values).
$\boldsymbol{W_e} \in R^{V\times d} $ is the embedding matrix. Each row in the matrix represents the embedding vector of a unique category. $d$ is the dimensionality of the embedding, which is a hyperparameter specified prior to training. $e_i \in R^{1\times d}$ is the embedding vector of every $x_i$ to which it belongs.

\noindent\textbf{Max-Mean and Bit-wise attention (MMBAttn) module.} In this part, all components of the designed attention mechanism including, max, mean and bit-wise attention will be explained in detail.
Accordingly, the embedding weights in the max and mean attention blocks can be calculated as follows

\begin{equation}
\mathcal{W}_{attn} = \sigma(f_{\{\boldsymbol{W_1}, \boldsymbol{W_2}\}}(g(e_{i}))) \in R^{1\times V},
\end{equation}

\noindent where $\sigma(.)$ is a Sigmoid function and $g(e_{i})$ is embedding-wise max and mean pooling and can be computed as $g_{max}(e_{i}) = \max_{i=1}^{d} e_i$ and $ g_{mean}(e_{i}) = \frac{1}{d}\sum^{d}_{i=1}{e_i}$, respectively, where $i \in [1, 2, ..., d ]$ and $s_i = g(e_{i})$ is a scalar value and it represents the global information about the $i-th$ embedding vector. $f_{\{\boldsymbol{W_1}, \boldsymbol{W_2}\}}$ is MLP and can be formulate as

\begin{equation}
f_{\{\boldsymbol{W_1}, \boldsymbol{W_2}\}} = \boldsymbol{W_2} \max(0, \boldsymbol{W_1}^T s_i),
\end{equation}

\noindent where $\boldsymbol{W_1} \in R^{{C/R}\times C}$ and $\boldsymbol{W_2} \in R^{C\times {C/R}}$ are MLP weights used to learn feature importance and activation function ReLU comes after $\boldsymbol{W_1}$. $R$ is reduction ratio. Output of max and mean attentions as

\begin{equation}
\mathcal{W}^{MM}_{attn} = \mathcal{W}^{\max}_{attn} + \mathcal{W}^{mean}_{attn} \in R^{1\times V},
\end{equation}

The re-weighted embedding vector obtained using max and mean attentions can be expressed as follows,
\begin{equation}
{F}^{MM} = e_i\otimes \mathcal{W}^{MM}_{attn} \in R^{1\times Vd},
\end{equation}

\noindent where $\otimes$ is an element-wise multiplication.

Similarly, the bit-wise embedding attention vector $\mathcal{W}^{B}_{attn}$ can be calculated as

\begin{equation}
\mathcal{W}^{B}_{attn} = \sigma(f_{\{\boldsymbol{W_1}, \boldsymbol{W_2}\}}(e_{cat})),
\end{equation}

\noindent where $e_{cat} = concat[e_1; e_2; e_3; ... e_d] \in R^{1\times Vd}$ is concatenated embedding vector.

\noindent The re-weighted embedding vector obtained using bit-wise attentions can be formulated by

\begin{equation}
{F}^{B} =\mathcal{W}^{MM}_{attn}\otimes \mathcal{W}^{B}_{attn} \in R^{1\times Vd},
\end{equation}

\noindent The final re-weighted embedding vector ${F}^{MMB}$ can be expressed as follows,

\begin{equation}
{F}^{MMB} = {F}^{MMB} + {F}^{MMB} \in R^{1\times Vd},
\end{equation}

\noindent Finally, this re-weighted embedding vector can be fed to any CTR prediction model.

\noindent\textbf{Training phase.} During the training phase, the widely used loss function for binary classification problems, also known as Binary Cross Entropy (BCE) or log loss, was employed. The formula for log loss is as follows:

\begin{equation}
log loss = -\frac{1}{N}\sum^{N}_{i=1}{(y\log(\hat{y}) + (1 - y)\log(1 - \hat{y}))}
\end{equation}

In this equation, $y\in$ \{0, 1\} represents the label value, $\hat{y} \in$ (0, 1) denotes the model's prediction, and $N$ denotes the total number of examples used during training.

\section{Experiments}

In this section, the datasets used in the experiments, the evaluation metrics for comparing the methods, and the neural network-based methods employed are discussed.

\subsection{Datasets}

\begin{table}[tbh]
\caption{The statistics of public datasets.}
\label{table:datasets}
\begin{center}
\begin{tabular}{|c c c c|} 
\hline
 Datasets & \# Instances & \# Fields & \# Features\\ [0.5ex] 
 \hline \hline
 Avazu & 40,428,967 & 22 & 1,544,250 \\
 \hline
  Criteo & 45,840,617 & 39 & 2,086,936 \\
 \hline
 Frappe & 288,609 & 10 & 5,382 \\  [0.25ex]   
 \hline
\end{tabular}
\end{center}
\end{table}

The experiments in this study were conducted on three publicly available datasets: Avazu, Criteo and Frappe. The statistical information for each dataset is listed in Table \ref{table:datasets}.

\textbf{Avazu\footnote{\url{https://www.kaggle.com/c/avazu-ctr-prediction/data}}}: Avazu dataset was published in a CTR prediction competition on Kaggle. It consists of chronologically ordered ad click data. The dataset includes an ID, click information, and 22 categorical features.

\textbf{Criteo\footnote{\url{http://labs.criteo.com/2013/12/download-terabyte-click-logs}}}: The Criteo dataset is widely used in many CTR modeling studies and is considered a benchmark for CTR prediction. The Criteo dataset comprises 26 anonymous categorical fields and 13 numerical feature fields.

\textbf{Frappe\footnote{\url{https://www.baltrunas.info/context-aware}}}: The Frappe dataset comprises a log of mobile app usage by users in different contexts. It includes entries for apps used by users in various situations. The target variable indicates whether the user has used a particular app in a specific context. This dataset comprises 10 fields.

\subsection{Evaluation Metrics}

Two evaluation metrics were used in the experiments. These metrics are AUC (Area Under ROC) and log loss (Cross entropy).

\textbf{AUC}: AUC represents the area under the ROC curve and is a commonly used evaluation metric for measuring the performance of classification problems. It measures the probability of assigning a higher score to a randomly chosen positive item than to a randomly chosen negative item for the CTR predictor. A higher AUC represents better performance.

\textbf{Log loss}: Log loss is a commonly used metric in binary classification problems. It measures the difference between the actual value and the predicted value. A smaller log loss value represents better performance.



\subsection{Implementation Details}
We implement our method based on the repository of FuxiCTR \cite{zhu2021open}, an open-source CTR prediction benchmark \footnote{\url{https://fuxictr.github.io}}. We experiment and report on reproducing the results of state-of-the-art models using the hyperparameters provided by this benchmark. To ensure the reliability of the results, we run the all base models 5 times. All the experiments are carried out with Python programming language (3.10.9) and Tensorflow (2.11.0) library on a system that has NVIDIA A40 GPU, AMD EPYC 7513 32-Core Processor, 100 GB RAM. 

\begin{table}[t]
\centering
\caption{Ablation studies for DNN model on Criteo and Avazu datasets.}
\label{table:performance}
\begin{tabular}{lcccccc}
\hline
\multirow{2}{*}{Model}                      & \multicolumn{3}{c}{Criteo} & \multicolumn{3}{c}{Avazu} \\
\cline{2-4} \cline{5-7}
                            & AUC    & Impr.   & LogLoss   & AUC     & Impr.    & LogLoss   \\
\hline \hline
DNN                         & 0.813  & Base    & 0.4381    & 0.7622  & Base    & 0.3681    \\
DNN + Mean                  & 0.8134 & 0.05\%  & 0.4381    & 0.7623  & 0.01\%  & 0.3683    \\
DNN + Max                   & 0.8136 & 0.07\%  & 0.438     & 0.76402 & 0.23\%  & 0.3673    \\
DNN + Bit-wise              & 0.8138 & 0.10\%  & 0.4379    & 0.76273 & 0.07\%  & 0.3681    \\
DNN + Max + Mean            & 0.8137 & 0.09\%  & 0.438     & 0.76489 & 0.34\%  & 0.3671    \\
DNN + Max + Mean + Bit-wise & 0.8143 & 0.15\%  & 0.4378    & 0.76499 & 0.37\%  & 0.3669    \\
\hline
\end{tabular}
\end{table}

\subsection{Baseline}
We compared our proposed model with several state-of-the-art models designed for CTR prediction and implemented in the FuxiCTR repository \cite{zhu2021open}. These models include Fibinet \cite{huang2019fibinet}, MaskNet \cite{wang2021masknet}, and AutoInt+ \cite{song2019autoint}, which all use attention mechanisms. Details about these models are provided in the introduction section. Although EDCN, DCN, and DeepFM \cite{deep&cross, chen2021enhancing, guo2017deepfm} do not utilize attention mechanisms, we included them in our comparison since they achieved the best results in the BarsCTR benchmark. Specifically, DCN and EDCN use MLP and Cross Networks with different fusion strategies, while DeepFM combines MLP and FM layers. FinalMLP \cite{mao2023finalmlp} is a two-stream model where each stream consists of MLP layers. One stream is fed with user-related features and the other with item-related features, and then a bilinear group aggregation strategy is used to fuse the two streams. The proposed attention mechanism has been applied to the baselines of DeepFM, FinalMLP, and EDCN models, as well as to a DNN model with 3 MLP layers of size 400 and ReLU activation function. This model is then compared to the same DNN model that uses the same hyperparameters but without attention.

\begin{figure}[ht]
\begin{center}
\includegraphics[width=\textwidth]{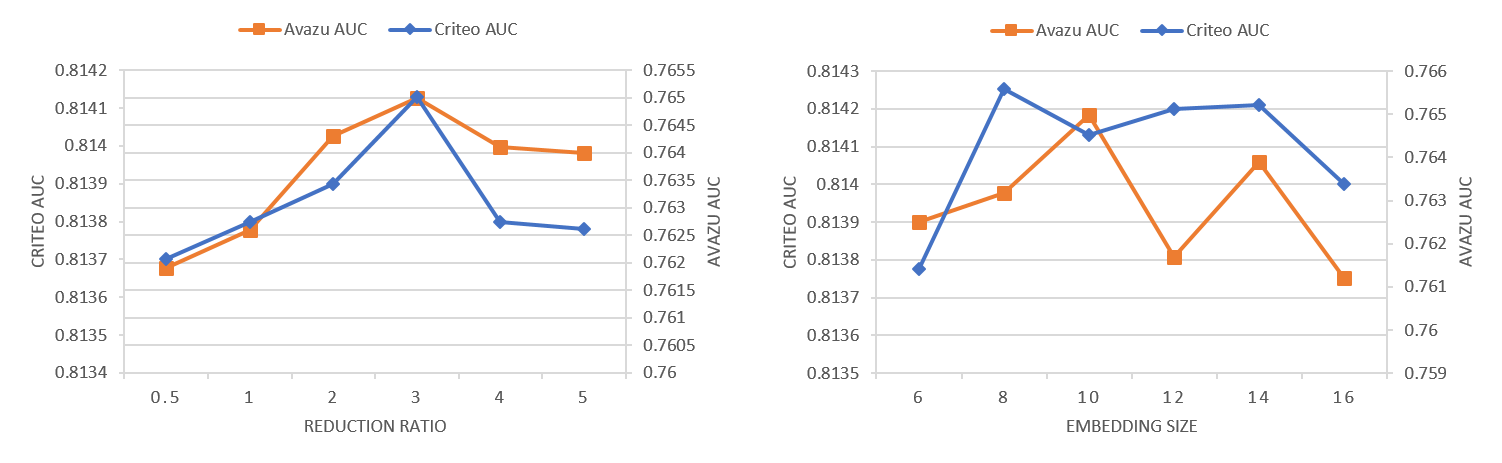}
\caption{Performance of ablations for different reduction ratios and embedding sizes on the Criteo and Avazu datasets.}
\label{fig:reduction}
\end{center}
\end{figure}

\subsection{Ablation Studies}
The results in Table ~\ref{table:performance} showcase the performance of all components of the proposed attention mechanism on the Criteo and Avazu datasets. The DNN model represents the baseline performance on both datasets. In both datasets, the max pooling attention mechanism significantly increased the AUC scores, while the mean pooling mechanism achieved comparable scores to the baseline. Especially in the Avazu dataset, max pooling improved the baseline model's result by 0.23\%. The bit-wise component improves the AUC scores by 0.1\% and 0.07\% on Criteo and Avazu datasets, respectively. When the maximum and mean pooling attention mechanisms are used together, it results in an improvement in the AUC scores for both datasets. Particularly, there is a significant improvement of approximately 0.34\% in the AUC score for the Avazu dataset. Additionally, the log loss value also decreases on this dataset.  When all the components are used together (MMBAttn), the highest improvement in AUC and log loss metrics is achieved. The proposed attention mechanism particularly provides a 0.37\% improvement in AUC for the Avazu dataset, reducing the log loss value from 0.3681 to 0.3669. This indicates that the combined use of maximum and mean pooling with the bit-wise attention mechanism effectively captures the intricate relationships between features and enhances the overall predictive performance of the CTR prediction model. The results demonstrate the effectiveness of the proposed approach in improving both accuracy and log loss metrics, especially on the Avazu dataset.
Fig. ~\ref{fig:reduction} displays an analysis of the reduction ratio and embedding size for the DNN + MMBAttn model on the Criteo and Avazu datasets. It is evident that both benchmarks have similar performance in terms of the AUC across various reduction ratios. Notably, the highest performances on both datasets are achieved with a reduction size of 3. Comparing the performance of the models for different embedding sizes, we can observe that embedding sizes 8 and 10 achieved the highest AUC values of 0.81425\% and 0.76499\% in the Criteo and Avazu datasets, respectively.

\subsection{State-of-the-art Comparison}

Table ~\ref{table:sota} shows the results of the proposed attention structure applied to DNN, DeepFM, EDCN and FinalMLP methods and its comparison with the results of other state-of-the-art methods. We used best hyperparameters obtained by ablation studies.
In all three datasets, the model with MMBAttn (Ours) improved model performances compared to the DNN baseline in terms of both AUC and Logloss metrics. Notably, the AUC is improved by 0.37\% on Avazu dataset while by 0.15\% on both Criteo and Frappe datasets. When we apply the proposed attention mechanism (MMBAttn) to the DeepFM model, the AUC scores are improved by 0.07\%, 0.092\% and 0.15\% on the Criteo, Avazu and Frappe datasets, respectively, while the log loss values decrease. Similarly, when we plug MMBAttn into the EDCN and FinalMLP models, the performance of the models showed a slight improvement compared to the DNN version model. Finally, several studies have shown that even a small improvement in terms of AUC metric in CTR prediction performance leads to large increases in online CTR scores and hence revenue \cite{wide&deep, ling2017ensemble, deep&cross}. Overall, even with the version of the MMBAttn mechanism plugged into the simple DNN version (DNN + MMBAttn) that does not utilized any complex interaction layers such as FM, CrossNet, Inner or Outer products etc., it achieved the second-best scores after the FinalMLP model which has state-of-the-art scores on all three datasets.

\begin{table}[ht]
\centering
\caption{State-of-the-art comparison on Criteo, Avazu and Frappe datasets.}
\label{table:sota}
\resizebox{\textwidth}{!}{
\begin{tabular}{lccccccccc}
\hline
\multirow{2}{*}{Model}                        & \multicolumn{3}{c}{Criteo}   & \multicolumn{3}{c}{Avazu}      &  \multicolumn{3}{c}{Frappe} \\
\cline{2-4} \cline{5-7} \cline{8-10}
                            & AUC    & Impr.   & LogLoss   & AUC     & Impr.   & LogLoss    & AUC     & Impr.   & LogLoss\\\hline \hline
Fibinet                     & 0.813  & -       & 0.4388    & 0.7645  & -       & 0.3673     & 0.9832  & -       & 0.1941\\
MaskNet                     & 0.8139 & -       & 0.438     & 0.7643  & -       & 0.3674     & 0.9837  & -       & 0.1696\\
DCN                         & 0.8138 & -       & 0.4381    & 0.7652  & -       & 0.3665     & 0.9839  & -       & 0.1527\\
AutoInt+                    & 0.8139 & -       & 0.4379    & 0.7645  & -       & 0.3668     & 0.9848  & -       & 0.149\\ \hline
DNN                         & 0.813  & Base    & 0.4381    & 0.7622  & Base    & 0.3681     & 0.9835  & Base    & 0.1605\\
DNN + MMBAttn (Ours)        & 0.8143 & 0.15\%  & 0.4378    & 0.765   & 0.37\%  & 0.3669     & 0.985   & 0.15\%  & 0.1441\\
DeepFM                      & 0.8137 & Base    & 0.4381    & 0.7648  & Base    & 0.3667     & 0.9842  & Base    & 0.1566\\
DeepFM + MMBAttn (Ours)     & 0.8143 & 0.07\%  & 0.4377    & 0.7655  & 0.092\% & 0.3666     & 0.9857  & 0.15\%  & 0.1507\\
EDCN                        & 0.8143 & Base    & 0.43766   & 0.7627  & Base    & 0.3681     & 0.9847  & Base    & 0.1576\\
EDCN + MMBAttn (Ours)       & 0.8145 & 0.02\%  & 0.43751   & 0.7637  & 0.13\%  & 0.3672     & 0.9852  & 0.051\% & 0.1476\\
FinalMLP                    & 0.81486 & Base   & 0.4372    & 0.7663  & Base    & 0.3659     & 0.9855  & Base    & 0.1487\\
FinalMLP + MMBAttn (Ours)   & 0.81497 & 0.01\% & 0.4371    & 0.7666  & 0.04\%  & 0.3657     & 0.9861  & 0.061\% & 0.1475\\
\hline
\end{tabular}}
\end{table}
\section{Conclusion}

In this work, we presented an attention-based approach, MMBAttn, for enhancing feature importance estimation in CTR prediction tasks in online advertising and recommendation systems. By leveraging max and mean pooling operations along with a bit-wise attention mechanism, the proposed method addresses the challenge of accurately determining feature importance, which is crucial for developing effective models in complex and large-scale CTR prediction scenarios. Traditional pooling operations, such as max and mean pooling, are widely used to extract relevant information from features. However, these operations can lead to information loss and hinder the accurate estimation of feature importance. To overcome this limitation, MMBAttn introduces a novel attention architecture that uses a bit-based attention structure that emphasizes the relationships between all bits in features, together with maximum and mean pooling. Extensive experiments were conducted on three public datasets to evaluate the effectiveness of the proposed method. The results demonstrate significant improvements in the performance of the base models, achieving state-of-the-art results. The proposed method is also designed as a plug-and-play module and is therefore model-agnostic, making it easily applicable to any CTR prediction model.
Future work can explore the application of the proposed attention mechanism in other domains and further investigate optimization techniques.

\bibliographystyle{unsrt}
\bibliography{references}

\begin{thebibliography}{10}

\bibitem{logistic2013trenches}
H.~Brendan McMahan, Gary Holt, D.~Sculley, Michael Young, Dietmar Ebner, Julian
  Grady, Lan Nie, Todd Phillips, Eugene Davydov, Daniel Golovin, Sharat
  Chikkerur, Dan Liu, Martin Wattenberg, Arnar~Mar Hrafnkelsson, Tom Boulos,
  and Jeremy Kubica.
\newblock Ad click prediction: A view from the trenches.
\newblock In {\em Proceedings of the 19th ACM SIGKDD International Conference
  on Knowledge Discovery and Data Mining}, KDD '13, page 1222–1230, 2013.

\bibitem{logistic2007}
Matthew Richardson, Ewa Dominowska, and Robert Ragno.
\newblock Predicting clicks: Estimating the click-through rate for new ads.
\newblock In {\em Proceedings of the 16th International Conference on World
  Wide Web}, WWW '07, page 521–530, New York, NY, USA, 2007. Association for
  Computing Machinery.

\bibitem{fm}
Steffen Rendle.
\newblock Factorization machines.
\newblock In {\em 2010 IEEE International Conference on Data Mining}, pages
  995--1000, 2010.

\bibitem{wide&deep}
Heng-Tze Cheng, Levent Koc, Jeremiah Harmsen, Tal Shaked, Tushar Chandra,
  Hrishi Aradhye, Glen Anderson, Greg Corrado, Wei Chai, Mustafa Ispir, Rohan
  Anil, Zakaria Haque, Lichan Hong, Vihan Jain, Xiaobing Liu, and Hemal Shah.
\newblock Wide \& deep learning for recommender systems, 2016.

\bibitem{guo2017deepfm}
Huifeng Guo, Ruiming Tang, Yunming Ye, Zhenguo Li, and Xiuqiang He.
\newblock Deepfm: A factorization-machine based neural network for ctr
  prediction, 2017.

\bibitem{deep&cross}
Ruoxi Wang, Bin Fu, Gang Fu, and Mingliang Wang.
\newblock Deep \& cross network for ad click predictions.
\newblock In {\em Proceedings of the ADKDD'17}, ADKDD'17, New York, NY, USA,
  2017. Association for Computing Machinery.

\bibitem{song2019autoint}
Weiping Song, Chence Shi, Zhiping Xiao, Zhijian Duan, Yewen Xu, Ming Zhang, and
  Jian Tang.
\newblock Autoint: Automatic feature interaction learning via self-attentive
  neural networks.
\newblock In {\em Proceedings of the 28th ACM International Conference on
  Information and Knowledge Management}, pages 1161--1170, 2019.

\bibitem{huang2019fibinet}
Tongwen Huang, Zhiqi Zhang, and Junlin Zhang.
\newblock Fibinet: combining feature importance and bilinear feature
  interaction for click-through rate prediction.
\newblock In {\em Proceedings of the 13th ACM Conference on Recommender
  Systems}, pages 169--177, 2019.

\bibitem{attn2021review}
Zhaoyang Niu, Guoqiang Zhong, and Hui Yu.
\newblock A review on the attention mechanism of deep learning.
\newblock {\em Neurocomputing}, 452:48--62, 2021.

\bibitem{zhang2022fibinet++}
Pengtao Zhang and Junlin Zhang.
\newblock Fibinet++: Improving fibinet by greatly reducing model size for ctr
  prediction.
\newblock {\em arXiv preprint arXiv:2209.05016}, 2022.

\bibitem{covington2016deep}
Paul Covington, Jay Adams, and Emre Sargin.
\newblock Deep neural networks for youtube recommendations.
\newblock In {\em Proceedings of the 10th ACM conference on recommender
  systems}, pages 191--198, 2016.

\bibitem{bahdanau2014neural}
Dzmitry Bahdanau, Kyunghyun Cho, and Yoshua Bengio.
\newblock Neural machine translation by jointly learning to align and
  translate.
\newblock {\em arXiv preprint arXiv:1409.0473}, 2014.

\bibitem{vaswani2017attention}
Ashish Vaswani, Noam Shazeer, Niki Parmar, Jakob Uszkoreit, Llion Jones,
  Aidan~N Gomez, {\L}ukasz Kaiser, and Illia Polosukhin.
\newblock Attention is all you need.
\newblock {\em Advances in neural information processing systems}, 30, 2017.

\bibitem{el2021automatic}
Wafaa~S El-Kassas, Cherif~R Salama, Ahmed~A Rafea, and Hoda~K Mohamed.
\newblock Automatic text summarization: A comprehensive survey.
\newblock {\em Expert systems with applications}, 165:113679, 2021.

\bibitem{yu2019deep}
Zhou Yu, Jun Yu, Yuhao Cui, Dacheng Tao, and Qi~Tian.
\newblock Deep modular co-attention networks for visual question answering.
\newblock In {\em Proceedings of the IEEE/CVF conference on computer vision and
  pattern recognition}, pages 6281--6290, 2019.

\bibitem{wang2020eca}
Qilong Wang, Banggu Wu, Pengfei Zhu, Peihua Li, Wangmeng Zuo, and Qinghua Hu.
\newblock Eca-net: Efficient channel attention for deep convolutional neural
  networks.
\newblock In {\em Proceedings of the IEEE/CVF conference on computer vision and
  pattern recognition}, pages 11534--11542, 2020.

\bibitem{hu2018squeeze}
Jie Hu, Li~Shen, and Gang Sun.
\newblock Squeeze-and-excitation networks.
\newblock In {\em Proceedings of the IEEE conference on computer vision and
  pattern recognition}, pages 7132--7141, 2018.

\bibitem{woo2018cbam}
Sanghyun Woo, Jongchan Park, Joon-Young Lee, and In~So Kweon.
\newblock Cbam: Convolutional block attention module.
\newblock In {\em Proceedings of the European conference on computer vision
  (ECCV)}, pages 3--19, 2018.

\bibitem{khan2022transformers}
Salman Khan, Muzammal Naseer, Munawar Hayat, Syed~Waqas Zamir, Fahad~Shahbaz
  Khan, and Mubarak Shah.
\newblock Transformers in vision: A survey.
\newblock {\em ACM computing surveys (CSUR)}, 54(10s):1--41, 2022.

\bibitem{dosovitskiy2021image}
Alexey Dosovitskiy, Lucas Beyer, Alexander Kolesnikov, Dirk Weissenborn,
  Xiaohua Zhai, Thomas Unterthiner, Mostafa Dehghani, Matthias Minderer, Georg
  Heigold, Sylvain Gelly, et~al.
\newblock An image is worth 16x16 words: Transformers for image recognition at
  scale.
\newblock {\em arXiv preprint arXiv:2010.11929}, 2020.

\bibitem{xiao2017attentional}
Jun Xiao, Hao Ye, Xiangnan He, Hanwang Zhang, Fei Wu, and Tat-Seng Chua.
\newblock Attentional factorization machines: Learning the weight of feature
  interactions via attention networks.
\newblock {\em arXiv preprint arXiv:1708.04617}, 2017.

\bibitem{zhou2019deep}
Guorui Zhou, Na~Mou, Ying Fan, Qi~Pi, Weijie Bian, Chang Zhou, Xiaoqiang Zhu,
  and Kun Gai.
\newblock Deep interest evolution network for click-through rate prediction.
\newblock In {\em Proceedings of the AAAI conference on artificial
  intelligence}, volume~33, pages 5941--5948, 2019.

\bibitem{feng2019deep}
Yufei Feng, Fuyu Lv, Weichen Shen, Menghan Wang, Fei Sun, Yu~Zhu, and Keping
  Yang.
\newblock Deep session interest network for click-through rate prediction.
\newblock {\em arXiv preprint arXiv:1905.06482}, 2019.

\bibitem{li2020interpretable}
Zeyu Li, Wei Cheng, Yang Chen, Haifeng Chen, and Wei Wang.
\newblock Interpretable click-through rate prediction through hierarchical
  attention.
\newblock In {\em Proceedings of the 13th International Conference on Web
  Search and Data Mining}, pages 313--321, 2020.

\bibitem{long2021efficient}
Chao Long, Yanmin Zhu, Haobing Liu, and Jiadi Yu.
\newblock Efficient feature interactions learning with gated attention
  transformer.
\newblock In {\em Web Information Systems Engineering--WISE 2021: 22nd
  International Conference on Web Information Systems Engineering, WISE 2021,
  Melbourne, VIC, Australia, October 26--29, 2021, Proceedings, Part II 22},
  pages 3--17. Springer, 2021.

\bibitem{zhang2019fat}
Junlin Zhang, Tongwen Huang, and Zhiqi Zhang.
\newblock Fat-deepffm: Field attentive deep field-aware factorization machine.
\newblock {\em arXiv preprint arXiv:1905.06336}, 2019.

\bibitem{juan2017field}
Yuchin Juan, Damien Lefortier, and Olivier Chapelle.
\newblock Field-aware factorization machines in a real-world online advertising
  system.
\newblock In {\em Proceedings of the 26th International Conference on World
  Wide Web Companion}, pages 680--688, 2017.

\bibitem{shi2021ctr}
Xiujin Shi, Yang Yang, and Chenyun Tao.
\newblock Ctr prediction model considering the importance of embedding vector.
\newblock In {\em 2021 IEEE International Conference on Artificial Intelligence
  and Computer Applications (ICAICA)}, pages 167--171. IEEE, 2021.

\bibitem{he2017neural}
Xiangnan He and Tat-Seng Chua.
\newblock Neural factorization machines for sparse predictive analytics.
\newblock In {\em Proceedings of the 40th International ACM SIGIR conference on
  Research and Development in Information Retrieval}, pages 355--364, 2017.

\bibitem{wang2021masknet}
Zhiqiang Wang, Qingyun She, and Junlin Zhang.
\newblock Masknet: Introducing feature-wise multiplication to ctr ranking
  models by instance-guided mask.
\newblock {\em arXiv preprint arXiv:2102.07619}, 2021.

\bibitem{zhu2021open}
Jieming Zhu, Jinyang Liu, Shuai Yang, Qi~Zhang, and Xiuqiang He.
\newblock Open benchmarking for click-through rate prediction.
\newblock In {\em Proceedings of the 30th ACM International Conference on
  Information \& Knowledge Management}, pages 2759--2769, 2021.

\bibitem{chen2021enhancing}
Bo~Chen, Yichao Wang, Zhirong Liu, Ruiming Tang, Wei Guo, Hongkun Zheng, Weiwei
  Yao, Muyu Zhang, and Xiuqiang He.
\newblock Enhancing explicit and implicit feature interactions via information
  sharing for parallel deep ctr models.
\newblock In {\em Proceedings of the 30th ACM international conference on
  information \& knowledge management}, pages 3757--3766, 2021.

\bibitem{mao2023finalmlp}
Kelong Mao, Jieming Zhu, Liangcai Su, Guohao Cai, Yuru Li, and Zhenhua Dong.
\newblock Finalmlp: An enhanced two-stream mlp model for ctr prediction.
\newblock {\em arXiv preprint arXiv:2304.00902}, 2023.

\bibitem{ling2017ensemble}
Xiaoliang Ling, Weiwei Deng, Chen Gu, Hucheng Zhou, Cui Li, and Feng Sun.
\newblock Model ensemble for click prediction in bing search ads.
\newblock In {\em Proceedings of the 26th International Conference on World
  Wide Web Companion}, WWW '17 Companion, page 689–698, Republic and Canton
  of Geneva, CHE, 2017. International World Wide Web Conferences Steering
  Committee.

\end{thebibliography}

\end{document}